\documentclass[pra,reprint,aps,superscriptaddress]{revtex4-1}
\usepackage[T1]{fontenc}
\usepackage{graphicx}
\usepackage{amsmath} 
\usepackage{amssymb}
\usepackage{amsfonts}
\usepackage{color}
\usepackage{comment}
\includecomment{comment}
\begin{document}
\title{Theory of relaxation oscillations in exciton-polariton condensates}
\author{Andrzej Opala}
\email{opala@ifpan.edu.pl}
\affiliation{Institute of Physics, Polish Academy of Sciences, Al. Lotnik\'ow 32/46, 02-668 Warsaw, Poland}
\author{Maciej Pieczarka}
\affiliation{OSN, Department of Experimental Physics, Faculty of Fundamental Problems of Technology, Wroc{\l}aw University of Science and Technology, W. Wyspia{\'n}skiego 27, 50-370 Wroc{\l}aw, Poland}
\author{Micha{\l} Matuszewski}
\affiliation{Institute of Physics, Polish Academy of Sciences, Al. Lotnik\'ow 32/46, 02-668 Warsaw, Poland}

\begin{abstract}
We provide an analytical and numerical description of relaxation oscillations in the nonresonantly pumped polariton condensate. The presented considerations are based on the open dissipative Gross-Pitaevskii equation coupled to a pair of rate equations. The evolution of the condensate density can be explained qualitatively by studying the topology of the trajectory in phase space. We use a fixed points analysis for the classification of the different regimes of condensate dynamics, including fast stabilization, slow oscillations and ultrashort pulse emission. We obtain an analytical condition for the occurrence of relaxation oscillations. Continuous and pulsed condensate excitation considered and we demonstrate that in the latter case the existence of the second reservoir is necessary for the emergence of oscillations. We show that relaxation oscillations should be expected to occur in systems with relatively short polariton lifetime.

\end{abstract}
\maketitle
\section{\label{sec:INTRO}INTRODUCTION}

Exciton-polaritons, coherently coupled excitons and photons in a semiconductor microcavity~\cite{Deng_1,Carusotto_1}, enabled the creation of a novel class of bosonic condensates characterized by non-equilibrium dissipative nature and complex nonlinear dynamics. The unique properties of polariton quantum fluids bring possibility to observe phenomena of superfluidity, instability, self-localization, self-trapped magnetic polarons, topological and nonlinear excitations~\cite{Ostrovskaya_1,Lagoudakis_1,Kartashov_1,Malpuech_1,Skryabin_1,Sich_1,Amo_1,Lerario_1,Hivet_1,soliton_dynamic,Flayac_1,Natalia_1,Michal_1,Pawel_1}. Description of polariton dynamics is an interesting physical problem important for the fundamental description of light-matter condensation 
and the potential polaritonic applications ~\cite{Hamid_1,Sanvitto_1,Fraser_1}. In this paper, we focus on the theoretical description of the nontrivial time evolution of polariton quantum fluid characterized by oscillatory behaviour. 


To date, oscillations of emission intensity in polariton systems have been investigated in several different contexts. The common phenomenon occuring in the case of resonant excitation is the Rabi oscillations between photonic and excitonic component of polaritons~\cite{Rabi_1,Weisbuch_1, Sanvitto_RabiOscillations}.
Spatial or spin oscillatory dynamics of exciton-polariton condensates were assigned in several works to the coupling of condensates forming an analog of a superconductor Josephson junction~\cite{Basia_1,Solnyshkov_1,Bloch_1}. Josephson oscillations take place between two macroscopic bosonic ensembles occupying single macroscopic states separated by tunnel barrier. The nontrivial polariton density oscillation in space was also studied in the context Zitterbewegung effect~\cite{Sedov}. 


Relaxation oscillations observed in polariton condensates have a completely different character than Rabi or Josephson oscillations, since there is no periodic transfer of particles between components constituting the polariton fluid. Instead, relaxation oscillations are due to periodic change in the efficiency of relaxation from the reservoir of uncondensed particles to the condensate, and may even take the form of sharp spikes, well separated in time. The first experimental observation of relaxation oscillation in polariton system was reported  in Ref.~\onlinecite{Milena_1}. The oscillatory behaviour was also studied in the context of ultrashort emission of pulses from polariton condensate propagating in a disordered potential~\cite{Pieczar_1}. 

From the general point of view, relaxation oscillations are a family of periodic solutions occurring in various dynamical systems. Mathematically, this type of dynamics can be observed in a certain class of coupled nonlinear differential equations. The most well known examples are the oscillating electrical triode circuits and B-class semiconductor lasers~\cite{Siegman_1, Shirley_1, Andronov_1}. These  systems can be described by a nonlinear Van der Pol equation. 
One of the characteristic properties of relaxation oscillations is the possibility of the presence of two stages in the cycle possessing different timescales~\cite{Grasman_1}. The first timescale is related to a slow change of phase and the second timescale represents rapid change due to a fast relaxation. Following this rapid change, the system can return to the stage of slow evolution~\cite{Grasman_1}.

Here, we study relaxation oscillations in an exciton-polariton condensate analyticaly and numerically using a model including both inactive and active reservoir coupled to the polariton condensate. We derive a second order differential equation for the condensate density, applicable in the regime of continuous and pulsed nonresonant excitation. We analyze the solutions to this equation in linear and nonlinear regimes, and demonstrate that oscillatory character of nonlinear fixed points explains the existence of relaxation oscillations in a certain parameter range. In the case of pulsed excitation, we find that the existence of the inactive reservoir is crucial for the appearance of oscillations. An analytical condition for the oscillatory regime is derived, which provides results consistent both with numerical simulations and with previous experimental observations. 

The paper is structured as follows. In Sec.~\ref{sec:MODEL}, we define the model based on the mean-field open-dissipative Gross-Pitaevskii and rate equations. 
In Sec.~\ref{sec:RESULTS} we investigate numerically the oscillatory evolution of the condensate. We describe in detail the mechanism of relaxation oscillations considering the stimulated scattering and evolution of the reservoirs. 
In Sec.~\ref{sec:FIXEDPOINT} we analyze fixed points of the model and the system time evolution in phase space. We derive an analytical condition for the existence of relaxation oscillations.
In Sec.~\ref{sec:CONCLUSION} we summarize our work. 

\section{\label{sec:MODEL}MODEL}
Modelling exciton-polariton condensates excited nonresonantly is a complex problem. In many cases, theoretical reconstruction of the experimental data must take into account processes of exciton reservoir formation, their relaxation and stimulated scattering to the condensate. In our work, we consider the evolution of two reservoirs called the active and inactive reservoir~\cite{Deveaud_VortexDynamics,Forchel_StepwiseDynamics}. The physical interpretation of these two reservoirs may differ in different experimental conditions. For example, in~\cite{Milena_1} the two reservoirs were attributed to excitons residing in the vicinity of the localized condensate and away from it. On the other hand, in~\cite{Sczytko_ExcitonFormation,Matuszewski_UniversalityPolaritons} the two components of dynamics were identified as the reservoir of excitons and free carriers (electron-hole plasma). Here, we do not specify the exact nature of the two fields, but assume that the system can be roughly divided into two sets of excitations and only the active reservoir modes provide direct stimulated scattering to the polariton condensate. Importantly, as we will show below, taking into account the second reservoir is necessary for the existence of oscillations in the case of pulsed excitation, which is relevant to recent experiments~\cite{Milena_1,Pieczar_1}.

The considered system can be modelled using the generalized open-dissipative Gross-Pitaevskii equation (ODGPE) and rate equations written respectively for both reservoirs. The ODGPE describes the temporal evolution of the complex polariton order parameter \(\Psi({\bf r},t)\).  This equation  is dynamically coupled to the rate equation for the active reservoir  \(n_R ({\bf r},t)\). We also include the equation describing population of the inactive reservoir \(n_I (t) \) generated by the nonresonant laser field.  
The equations take the form presented below
\begin{equation}\label{1}
i \hbar \frac{\partial\Psi} {\partial t} = \left[-\frac{{\hbar}^2}{2m^{*}}\nabla^2+\frac{i \hbar}{2}(R n_R -\gamma_C)+U\right]\Psi
\end{equation}
\begin{equation}\label{2}
\frac{\partial n_R} {\partial t} = \kappa^{1D}n_I^2-\gamma _Rn_R-Rn_R|\Psi|^2
\end{equation}
\begin{equation}\label{3}
\frac{\partial n_I} {\partial t} = P-\kappa^{1D}n_I^2-\gamma _I n_I
\end{equation}
where: \(m_{LP}^*\) is the effective mass of lower polaritons and \(P({\bf r},t)\) is the laser pumping rate. The scattering into the reservoir and stimulated scattering rate into the condensate are described by \(\kappa^{1D}\) and \(R\). 
We assume that the scattering from the inactive reservoir and the stimulated relaxation of polaritons are respectively given by the function \(\kappa^{1D} n_I({\bf r},t)^2\)
and the function \(n_R({\bf r},t) R\). While the form of scattering terms may be different for a particular physical interpretation, the results presented in this work would not change qualitatively. The losses in the system are characterized by the parameters \(\gamma_I\), \(\gamma_R\) and \(\gamma_C\) describing the decay of inactive and active reservoirs and decay of polaritons, respectively. The decay rate of polaritons \(\gamma_C\) incorporates the finite lifetime of the photon component in semiconductor microcavity and the none-radiative decay rate of excitons. The potential \(U({\bf r},t)\) is the effective potential composed of the static potential due to sample design or disorder, the mean-field interaction within the condensate and between the condensate and reservoirs.

In our work, we will limit our considerations to the case when the condensate is practically trapped in a potential well, in which case it is possible to use simplified single-mode approximation to the condensate dynamics. We will use the same approximation to the reservoir fields, although allow that the active and inactive reservoir may posses different spatial distributions than the condensate, as suggested in~\cite{Milena_1,Stepnicki_TightBinding}. Nevertheless, we will neglect any changes of spatial distributions during the oscillations, which however still allows us to describe the main physics behind these phenomena. The reduced set of equations is the starting point of our work, and reads
\begin{align}
\label{GP_single}
\frac{d n_C} {d t} &= R n_R n_C -\gamma_C n_C, \\
\frac{d n_R} {d t} &= \kappa^{1D}n_I^2 - \gamma _R n_R - Rn_R n_C, \nonumber\\
\frac{d n_I} {d t} &= P(t) - \kappa^{1D} n_I^2 - \gamma _I n_I. \nonumber
\end{align}
In our considerations we will consider in detail two limiting cases, the continuous wave pumping where $P(t)={\rm const}$, and ultrashort pulse pumping where $P(t)=P_0\delta(t)$.





\section{Results} \label{sec:RESULTS}
For clarity, we begin our presentation with an example of relaxation oscillations in the case of continuous wave excitation. We solved the equations (\ref{1}{-}\ref{3}) numerically with a  4-th order Runge-Kutta method. Exemplary results are presented in~Fig.~1. The  details of simulation parameters are given in Table~1 in the Appendix.  To obtain complex and relevant polariton condensate dynamics, we used parameter values similar to the ones used in the experimental works~\cite{Milena_1,Pieczar_1}. In particular, we chose a short polariton lifetime with respect to the exciton reservoir decay rate, which was inspired by the observation that the oscillating dynamics appears naturally in lower quality microcavities \cite{Pieczar_1,Milena_1}. Initially, the reservoir fields are assumed to be empty and the condensate density is seeded with a small nonzero value, so that condensate buildup can occur according to Eq.~(\ref{GP_single}). We assume that such a small seed may result from thermal or quantum fluctuations in the polariton field, but the investigation of its origins is beyond this work. We emphasize that the results do not depend qualitatively on the value of the initial seed. 

\begin{figure}
\includegraphics[width=0.40\textwidth]{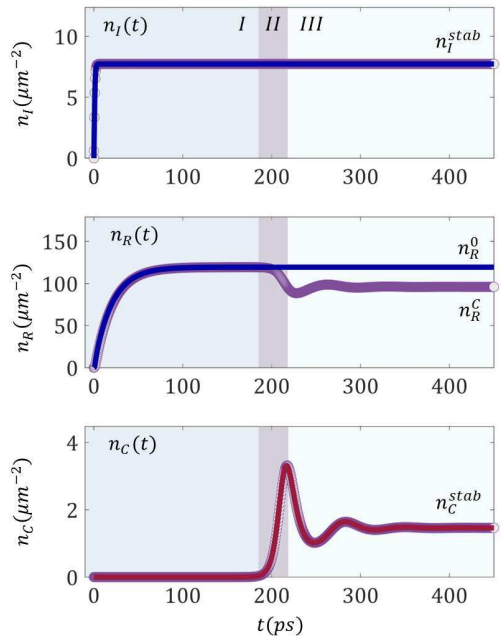}
\caption{Time evolution of the polariton population \(n_C(t)\) and incoherent reservoirs \(n_I(t)\) nad \(n_R(t)\) generated by continuous wave excitation. The blue  line corresponds to the analytical solution describing the evolution of reservoirs in the assumption of the negligible polariton field in the system. The red line present condensate evolution obtained by nonlinear oscillator aproximation. The grey points correspond to the numerical solution obtained by a solved full form of the open-dissipative Gross-Pitaevski equation.   The condensate parameters are presented in the Table~1.}
\label{fig:Fig_1}
\end{figure}

As shown in Fig.~\ref{fig:Fig_1}, initially the non-resonant pumping generates the population in the inactive reservoir $n_I$, which builds up quickly and saturates at a stationary level. 
The long-lived inactive reservoir relaxes and feeds the active reservoir. The active reservoir density is increased up to a level determined by the stationary value $n_R^0$.
The condensate density is negligibly small. This phase of dynamics is marked as stage I in~Fig.~\ref{fig:Fig_1}. 

Next, the increasing density of the exciton reservoir results in accumulation of the polariton condensate density due to the stimulated relaxation process. The polariton field grows exponentially. This short phase of condensate dynamics is presented as stage II in Fig.~\ref{fig:Fig_1}.

The relaxation oscillations in the condensate develop when the rapid condensate density growth stops due to the resulting depletion of the active reservoir. 
In this stage III in Fig.~\ref{fig:Fig_1}, the system is described by two timescales. The first timescale is related to the fast oscillations and the condensate decay rate. The second timescale is slower and it is related to the active reservoir decay. In each cycle of oscillations, the active reservoir density is initially depleted by the strong scattering to the condensate. The decreased reservoir density cannot sustain the condensate, which leads to fast decay of polariton field, due to its large decay rate $\gamma_C$. In the second stage of the cycle, The reduced stimulated scattering allows the active reservoir to be replenished. In effect, the minima and maxima of reservoir and condensate density are shifted in time with respect to each other by one-quarter of a cycle, see Fig.~\ref{fig:Fig_1}. These oscillations repeat until the condensate and reservoir reach the equilibrium level.


\subsection{\label{sec:RESERVOIR_DYNAMIC} Dynamics of reservoirs}
In order to investigate quantitatively the oscillatory behaviour of the polariton condensate, we determine the temporary dynamics of the inactive  (\ref{2}) and active reservoir (\ref{3}). 
In the case of continuous wave condensate excitation \(P(t)=\) const. The solution of the equation (\ref{3}) takes the form 
\begin{equation}\label{6}
n_I(t)=\frac{1}{2}\frac{\tanh{\Big(\frac{1}{2}t\sqrt{\eta^2+\gamma_I^2+\xi^0}\Big)\sqrt{\eta^2+\gamma_I^2}-\gamma_I}}{\kappa^{1D}},
\end{equation}
where \(\eta=\sqrt{4P_0\kappa^{1D}}\) and \(P_0=\alpha P_{th}\). The parameter
\begin{equation}\label{7}
\xi^0=\tanh^{-1}\Bigg(\frac{2n^{in}_I\kappa^{1D}+\gamma_I}{\sqrt{\eta^2+\gamma_I^2}}\Bigg),
\end{equation}
 takes into account the initial level of inactive reservoir density \(n_I^{in}\). When this reservoir is empty initially, i.e. when we consider turning on the pump abruptly with \((n_I(0)=n_I^{in}=0)\) and \(\gamma_I \ll \gamma_C,\gamma_R\), the parameter \(\xi^0\) is negligibly small. The temporal evolution of $n_I(t)$ according to~(\ref{6})  is presented in Fig.~1 (top) with a blue solid line. 

Typically, the population of the inactive reservoir given by equation (\ref{3}) stabilizes relatively quickly after turning on the pump at $t=0$. This is due to the asymptotic nature of the hyperbolic tangent solution of (\ref{7}). The steady state level of inactive reservoir density is obtained by eqating the left hand side of the equation (\ref{3}) to zero, which results in 
\begin{equation}\label{8}
n_I^{stb}=\frac{-\gamma_I+\sqrt{4\kappa^{1D}P_0+\gamma_I^2}}{2 \kappa^{1D}}.
\end{equation}
Stabilization of  the inactive reservoir in stage I allows to use the assumption that the inactive reservoir is constant during the subsequent stages (II, III) of evolution.

Next, we consider the dynamics of the active exciton reservoir $n_R$. We assume the absence of excitons in the active reservoir \(n_R(0)=0\) at the initial time.  We consider two cases, that is in the presence and absence of the condensate. These two types of dynamics have different solutions which also depend on the condensate density. First, we assume no polaritons in the condensate \(|\psi(t)|^2=0\). This assumption is correct below threshold intensity  when \((n_I^{stb})^2\kappa^{1D}<P_{th}=\gamma_C \gamma_R/R\), or above threshold at the initial stage of the system evolution (stage I), when polariton density is still negligibly small in comparison to the reservoir density. The active reservoir equation (\ref{GP_single}) has the analytic solution
\begin{equation}\label{9}
n_R(t)=\frac{(n_I^{stb})^2\kappa^{1D}}{\gamma_R}(e^{-t}-1),
\end{equation}
where the characteristic level of the stationary density is
\begin{equation}\label{10}
n_R^{0}=\frac{P_I}{\gamma_R},
\end{equation}
where \(P_I=(n_I^{stb})^2\kappa^{1D}\) describes the effective pumping intensity generated by the inactive reservoir. 
On the other hand, as follows from Eq.~(\ref{GP_single}), above threshold \(P_I>P_{th}\) when the condensate density achieves the stationary level, the reservoir density is equal to
\begin{equation}\label{11}
n_R^{C}=\frac{\gamma_C}{R}.
\end{equation}
We can introduce the \(\Delta n\) parameter describing the difference between reservoirs level in the case of the condensate absence \(n_R^0\)  and presence \(n_R^C\)
\begin{equation}\label{12}
\Delta n=\frac{P_I-P_{th}}{\gamma_R}.
\end{equation}
The value of this parameter can be connected to the amplitude of the oscillations of the active reservoir. 
In the case of large \(\Delta n\) stabilization of the condensate density is accompanied by noticeable oscillations.

In the case of excitation of the system with an ultrashort optical pulse, 
we assume that time evolution starts just after the arrival of the pulse. The pulse generates a certain density of the inactive reservoir \(n_I (0)=n_I^{in}\) at \(t=0\). The relaxation and decay of the inactive reservoir density is described by
\begin{equation}\label{13}
\frac{d}{d t} n_I(t)=-\kappa^{1D}n_I(t)^2-\gamma_I n_I(t).
\end{equation}
The above equation have the analytical solution
\begin{equation}\label{12}
n_I(t)=\frac{\gamma_I n_I^{in}}{(\kappa^{1D}n_I^{in}+\gamma_I)\exp(\gamma_I t )-\kappa^{1D}n_I^{in}}.
\end{equation}
This equation describes the quasi-exponential decay of the inactive reservoir population. Note that it is independent of the active reservoir and condensate density.

 \begin{figure}
\includegraphics[width=0.45\textwidth]{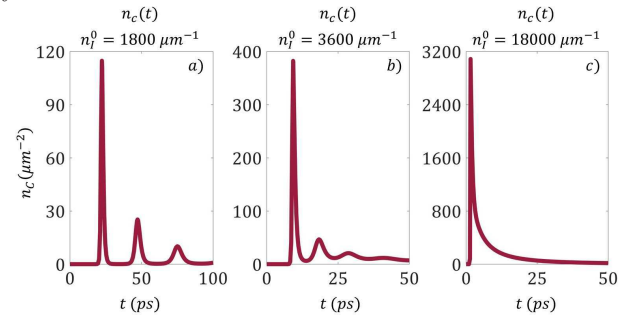}
\caption{Evolution of polariton condensate density under ultrashort pulsed excitation. The above three panels show respectively (a) spiked pulse emission, (b) decaying harmonic-like oscillations and (c) smooth stabilization. Simulation parameters are given in Table~1.}
\label{fig:pulsed}
\end{figure}

\subsection{\label{sec:CONDENSATE_DYNAMIC} Condensate Dynamics}

In this subsection we analyze the full dynamics of the system including the polariton condensate. We present nontrivial oscillatory dynamics resulting from the coupling between the active reservoir and the condensate. 



Our starting point is the set of evolution equations in the single mode approximation, Eqs.~(\ref{GP_single}). To demonstrate the qualitative properties of its solutions, we note that the inactive reservoir is not influenced by the other two components. Moreover, its dynamics in the oscillatory stage are typically slow, both in the case of pulsed and continuous wave excitation. This is due to the small decay rate $\gamma_I$, which physically is related the low probability of radiative recombination of high energy excitations. Therefore, for qualitative analysis it is reasonable to approximate the inactive reservoir density by a constant value.
In the case of continuos wave excitation, it is equal to the stationary value given by equation (\ref{8}), while in the pulsed excitation case it is the instantaneous value of $n_I$ which is assumed to decay very slowly. We combine the two remaining equations of the set~(\ref{GP_single}) for $n_C$ and $n_R$ to obtain the second order equation for the evolution of polariton density 
\begin{widetext}
  \begin{equation}\label{122}
    \frac{d^2 n_C} {d t^2} =-(\gamma_C\gamma_R + \gamma_CR n_C(t)-P_IR)n_C-(Rn_C(t)+\gamma_R)\frac{d n_C} {d t}+\frac{\left(\frac{d }{d t} n_C (t)\right)^2}{n_C (t)}.
  \end{equation}
\end{widetext}
where $P_I=n_I^2\kappa^{1D}$. The above form of the equation suggests that the solutions may have the form similar to those of a damped harmonic oscillator. However, we note that this is not the case in the low polariton density limit. Indeed, it is straightforward to show that in the linear limit of $n_C \rightarrow 0$ and with the substitution $x=\dot{n}_C/n_C$ the above equation reduces to
\begin{equation}\label{Eqx}
    \dot{x}=-Ax-B,
\end{equation}
where $A$ and $B$ are constants, which has an exponentially decaying solution. Therefore any oscillatory behaviour may only occur in the nonlinear regime, where the terms second order in $n_C$ in Eq.~(\ref{122}) provide qualitative corrections to the dynamics. The more detailed mathematical analysis of the nonlinear regime is presented in the next subsection.

\begin{figure}
\includegraphics[width=0.41\textwidth]{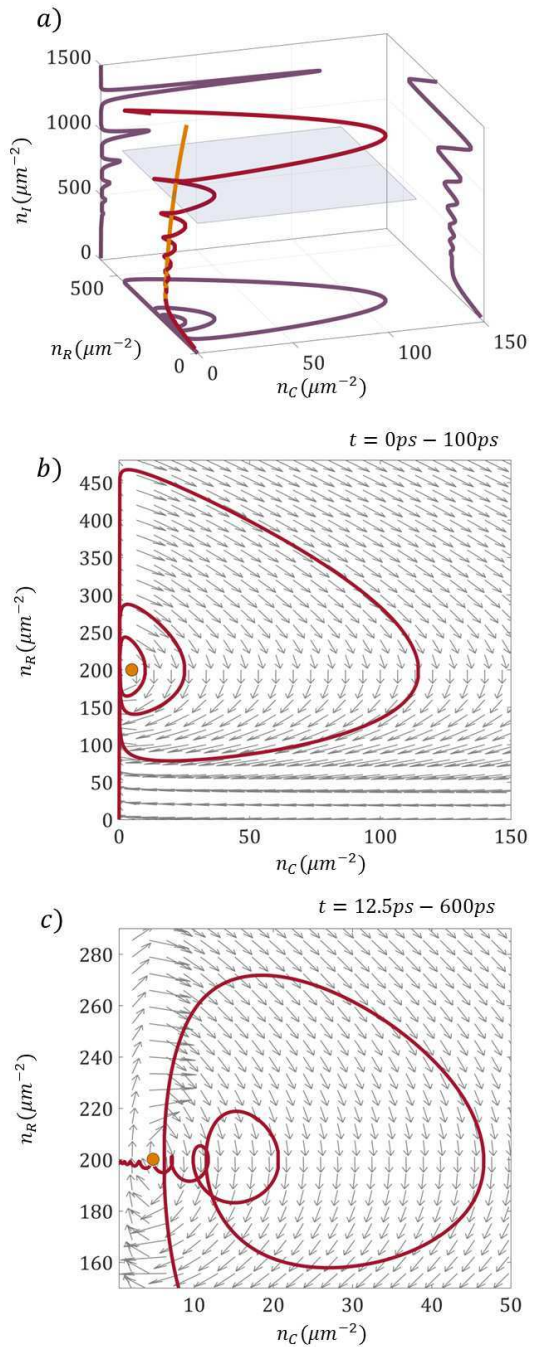}
\caption{(a) Phase space of trajectory in the pulsed excitation case of Fig.~\ref{fig:pulsed}(a) (red line). Purple lines correspond to the evolution projected respectively to \((n_I,n_C)\), \((n_I,n_R)\) and \((n_R,n_C)\) subspaces. The spiral evolution corresponds to the oscillatory behaviour of the system, with a gradually decaying inactive reservoir density $n_I$. 
It follows the evolution of the stationary state at a particular value of $n_I$  (orange line). 
(b) Cross-section of the phase space corresponding to the blue surface in panel (a). The arrows indicate the gradient of the evolution in the ($n_C$, $n_R$) subspace at the time when the value of \(n_I\) is equal to \(1000\mu m^{-1}\). The orange dot is the center of the spiral which is a stationary point within this subspace. (c) Same as (b) but in the case of decaying harmonic-like oscillations of Fig.~\ref{fig:pulsed}(b).}
\label{fig:phasespace}
\end{figure}

To demonstrate the character of complex oscillatory dynamics of the system we present the evolution described by the full model~(\ref{GP_single}). The typical evolution of condensate density is presented in Fig.~\ref{fig:pulsed}, showing (a) the spiked pulsed emission, (b) decaying oscillations, and (c) quasi-exponential decay. Note that all these three behaviors were observed in experiments~\cite{Milena_1,Pieczar_1}. The evolution in a three-dimensional phase space in the case of Fig.~\ref{fig:pulsed}(a) is presented in  Fig.~\ref{fig:phasespace}(a).
The spiral corresponds to the periodic behaviour of the system, with a gradually decaying inactive reservoir density $n_I$. 
In panel (b) in Fig.~\ref{fig:phasespace} we present a cross-section of phase space corresponding to the blue surface in panel (a). The arrows correspond to the gradient of the evolution in the ($n_C$, $n_R$) subspace at the time when the value of \(n_I\) is equal to the \(1000\mu m^{-1}\). The orange dot is the centre of the spiral which is a stationary point within the subspace.
In panel (c) a cross-section of phase space is presented in the case of spiked pulsating dynamics of Fig.~\ref{fig:pulsed}(a). In this case, instead of vibrations around a stationary path in phase space, large-amplitude pulsations occur as the condensate density intermittently decreases to zero. The phase space evolution is characterized by a spiral flattened at the $n_C=0$ axis.
{\color{blue}}

The remarkable characteristic of the oscillatory dynamics 
is the change of the character of the solution type from exponential to oscillatory in the subsequent stages of the evolution. 
These qualitative changes in the dynamics are related to bifurcations and are analyzed in the next section. 

\section{\label{sec:FIXEDPOINT} Fixed point analysis}

 \begin{figure}
\includegraphics[width=0.45\textwidth]{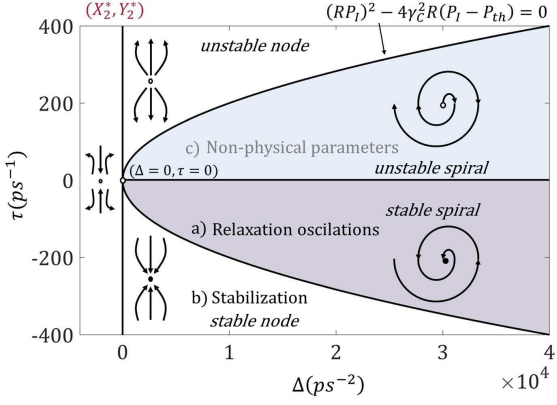}
\caption{Schematic presentation of the types of fixed points. The figure axes correspond to the trace \(\tau\) and the determinant \(\Delta \) of the Jacobian A. The bold black line corresponds to the analytical condition for relaxation oscillations.}
\label{fig:schematic}
\end{figure}

In this subsection we analyze the fixed points of Eq.~(\ref{122}) and determine the character of solutions around these points in the approximation of small perturbations. Such analysis is well suited for the description of solutions of nonlinear differential equations.
These investigation adapted to the obtained nonlinear polariton equation allows to determine the analytical condition for oscillations by classification of fixed points which generally can be a spiral, a centre, a saddle or a stable/unstable node~\cite{Strogatz_1}. 

We rewrite Eq.~(\ref{122}) in the form
\begin{equation}\label{21a}
 \begin{cases} \dot{X}  = f(Y) \\  \dot{Y}=g(X,Y) \end{cases} 
\end{equation}
where \(X=n_C(t)\) and \(Y=\frac{d}{d t} n_C(t)\), which results in
\begin{equation}\label{21}
 \begin{cases} \dot{X}  = Y, \\ 
 \dot{Y}=-(\gamma_C\gamma_R + \gamma_CR X-{P_I} R)X\\-(RX+\gamma_R)Y+\frac{Y^2}{X}. \end{cases} 
\end{equation}
We consider the phase portrait in vicinity of fixed points \((X^*,Y^*)\) which fulfill the conditions \(f(X^*,Y^*)=0\) and \(g(X^*,Y^*)=0\). 

Equation~(\ref{21}) is characterized by two fixed points \((X_1^*,Y_1^*)\) and \((X_2^*,Y_2^*)\). The first point  \(Y_1^*=0\), \(X_1^*=0\) corresponds to the absence of condensate in the system. As we demonstrated in the previous section, in the vicinity of this point, i.e.~in the limit $n_C\rightarrow 0$, the system cannot display any oscillatory dynamics as the evolution has the character of a smooth decay described by Eq.~(\ref{Eqx}).

An important consequence of the absence of oscillations around \((X_1^*,Y_1^*)\) is that they cannot appear in the model with only one reservoir, in the case of pulsed excitation. Indeed, considering the following simplified model
\begin{align}
\label{GP_1reservoir}
\frac{d n_C} {d t} &= R n_R n_C -\gamma_C n_C, \\
\frac{d n_R} {d t} &= P(t) - \gamma _R n_R - Rn_R n_C, \nonumber
\end{align}
where we removed the inactive reservoir and replaced the scattering term with direct pumping $P(t)$. After arrival of the pumping pulse this system would follow the evolution given by the above equation with $P(t>0)=0$, which does not possess the nontrivial fixed point \((X_2^*,Y_2^*)\). Therefore, the second, inactive reservoir is crucial for the existence of relaxation oscillations in the pulsed excitation case. Note that in the continuous wave excitation case the second reservoir is not crucial for the existence of oscillations since the continuous external pumping would have the same effect as the nonzero $P_I$ term in the full set of equations.
The second fixed point \(Y_2^*=0\) and \(X_2^*=(P_I-P_{th})/(\gamma_C)\) corresponds to the above-threshold $(P_I>P_{th})$ stationary solution achieved when the condensate is stabilized, and it will be considered below.
We can classify the different types of system dynamics using linearization around \((X_2^*,Y_2^*)\). This technique can reveal the phase portrait near the fixed point and determined their type~\cite{Strogatz_1}.
We introduce a small perturbation given by \(u=X-X^*\) and \(v=Y-Y^*\), with \(\dot{u}=\dot{X}\) and \(\dot{v}=\dot{Y}\). Using Taylor series expansion we rewrite (\ref{21}) in the form 
\begin{equation}\label{22}
 \begin{cases} \dot{u}  = u\frac{\partial f}{\partial X} + v\frac{\partial f}{\partial Y} + O(u^2,v^2,uv) \\  
 \dot{v} = u\frac{\partial g}{\partial X} + v\frac{\partial g}{\partial Y} + O(u^2,v^2,uv) \end{cases} 
\end{equation}
where the derivatives are taken at \((X_2^*,Y_2^*)\).
The perturbation \((u,v)\) evolves as
\begin{equation}\label{23}
\binom{\dot{u}}{\dot{v}}\approx\mathbf{A}\binom{u}{v}
\end{equation}
where we neglected second order terms and \(\mathbf{A}\) is the Jacobian matrix 
\begin{equation}\label{24}
\mathbf{A} = \left( \begin{array}{cc}
\frac{\partial f}{\partial X} & \frac{\partial f}{\partial Y}  \\
\frac{\partial g}{\partial X} & \frac{\partial g}{\partial Y}  \end{array} \right)_{|(X^*,Y^*)}
\end{equation}

We find  the eigenvalues of the problem \(\lambda\)
\begin{equation}\label{25}
\lambda^{I,II}_{(X^*,Y^*)}=\frac{\tau_{(X^*,Y^*)} \pm \sqrt{\tau_{(X^*,Y^*)}^2-4\Delta_{(X^*,Y^*)}}}{2}.
\end{equation}
where $\tau$ and $\Delta$ are the trace and the determinant of the Jacobian, respectively, given by
\begin{equation}\label{26}
\tau=-RX^*-\gamma_R+\frac{2Y^*}{X^*},
\end{equation}
\begin{equation}\label{27}
\Delta=-R P_I+2\gamma_C R X^* +\gamma_C\gamma_R +RY^*+\frac{{Y^*}^2}{{X^*}^2}.
\end{equation}

The solution of equation (\ref{25}) can be real or complex. The full diagram of possible fixed points is presented in Fig.~\ref{fig:schematic}. 
In the case when the eigenvalues are complex, the fixed point is a stable spiral which attracts trajectories surrounding it, while they perform oscillations. Such trajectories are similar to damped harmonic oscillations, and in our case spirals correspond to relaxation oscillations.  
On the other hand, when the eigenvalues are real, the evolution is a simple exponential decay and there are no oscillations.

Substituting Eqs.~(\ref{26}) and~(\ref{27}) into~(\ref{25}) and substituting the analytical form of \((X_2^*,Y_2^*)\), we obtain the analytical condition for relaxation oscillations given by the square root of \(\tau_{(X^*_2,Y^*_2)}^2-4\Delta_{(X^*_2,Y^*_2)}\)
\begin{equation}\label{28}
(RP_I)^2-4\gamma_C^2R(P_I-P_{th})>0,
\end{equation}
which determines whether the fixed point is a stable spiral or a stable node.



 \begin{figure}
\includegraphics[width=0.45\textwidth]{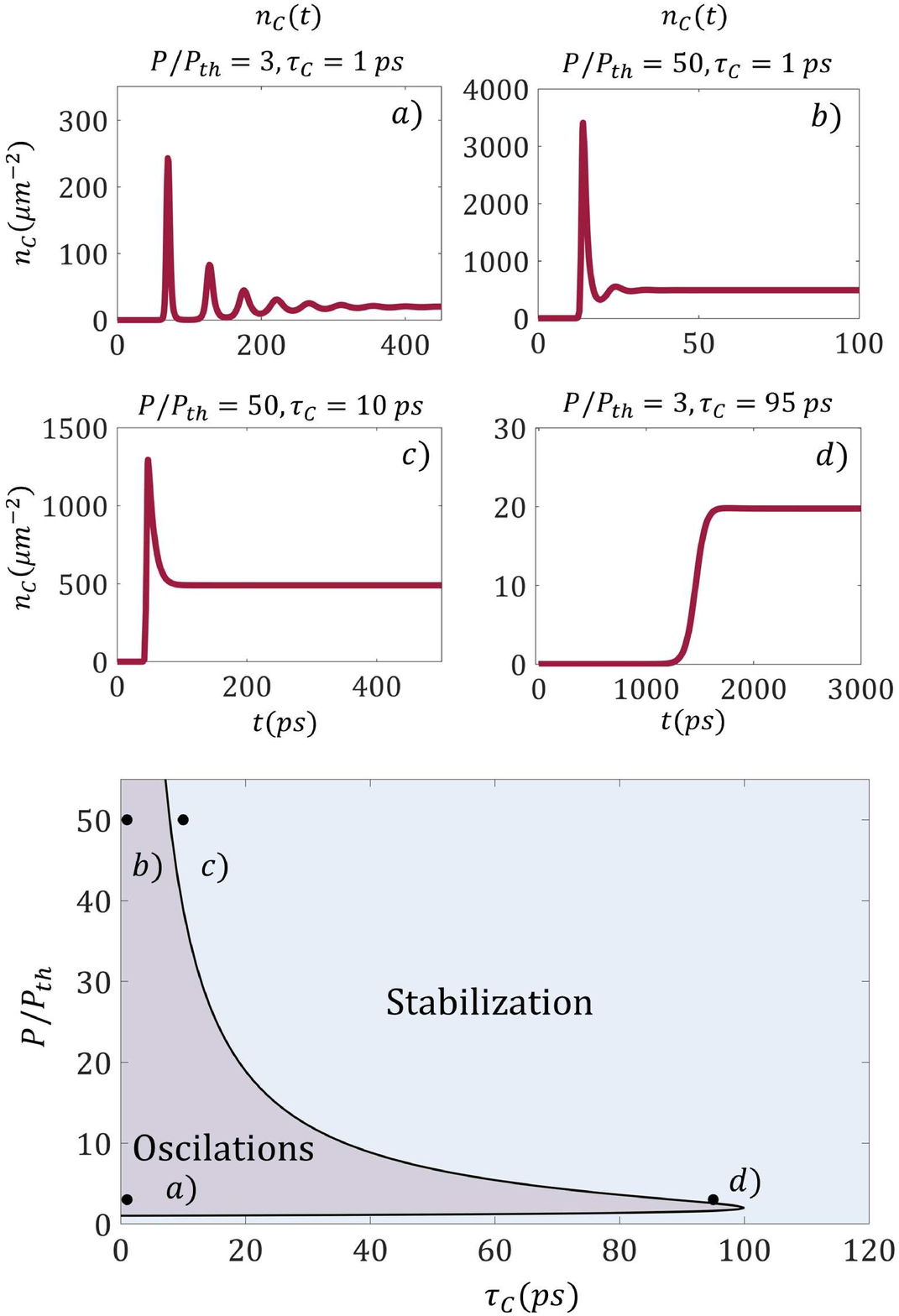}
\caption{(Bottom) Diagram showing region in parameter space which exhibits oscillations in the case of continuous wave excitation, according to the analytical formula, Eq.~(\ref{28}). (a)-(d) Examples of evolutions of condensate density corresponding to the points marked in the diagram.}
\label{fig:fig.5}
\end{figure}

 \begin{figure}
\includegraphics[width=0.50\textwidth]{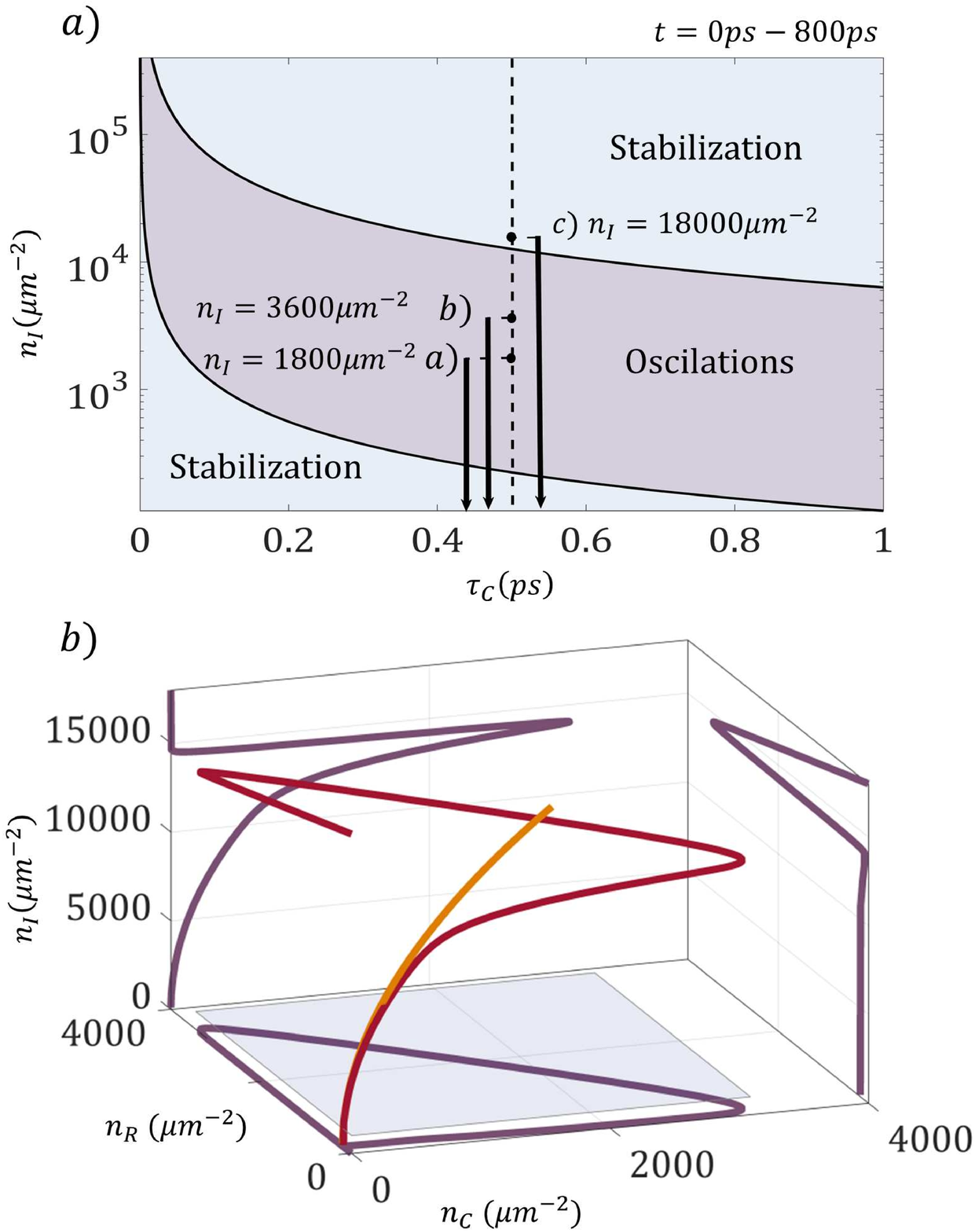}
\caption{(a) Diagram similar as in Fig.~\ref{fig:fig.5}, but in the case of pulsed excitation. Here, instead of power $P/P_{th}$, we plot the inactive reservoir density, which decays slowly during the evolution. The arrows indicate the paths in parameter space corresponding to the panels in Fig.~\ref{fig:pulsed}. While all evolutions traverse the oscillatory region, in the case (a) there are no visible oscillations because the system "sticks" to the stationary state while passing through it, as shown in panel (b).}
\label{fig:fig.6}
\end{figure}

The analytical condition for relaxation oscillations~(\ref{28}) and the equation for the inactive reservoir stationary density (\ref{8}) allow for the determination of a  diagram in parameter space, presented in Fig.~\ref{fig:fig.5}, together with typical system dynamics. For continuous wave excitation and fixed parameters $R$ and $\gamma_R$, the region in parameters space corresponding to oscillations is broader for a short polariton lifetime, which means that this phenomenon should be pronounced for relatively low-quality samples with short polariton lifetimes, see Fig.~\ref{fig:fig.5}(b). Nevertheless, even for long polariton lifetime oscillations can be observed at low pumping. The oscillations can have the character of spiked pulsations as in Fig.~\ref{fig:fig.5}(a) and at higher pumping power quickly decaying oscillations as in Fig.~\ref{fig:fig.5}(b). 
These observations are in qualitative agreement with the experimental data presented in~\cite{Milena_1}.

In the case of pulsed excitation, presented in Fig.~\ref{fig:fig.6}, the character of oscillations can be determined only for a given value of $n_I(t)$, which is assumed to change slowly in time. Consequently, during the evolution, the system may evolve from exponential decay, through oscillatory, and back to exponential stage. This is illustrated in panel (a) where the paths of evolution corresponding to Fig.~\ref{fig:pulsed} are marked. While all the paths go through the oscillatory region, the evolutions in Figs.~\ref{fig:pulsed}~(a)-(c) are markedly different. In particular, in Fig.~\ref{fig:pulsed}~(c) there are no visible oscillations. This behaviour can be explained by analysis of the trajectory in the full phase space in Fig.~\ref{fig:fig.6}~(b). Before reaching the oscillatory region, the system "sticks" to the stationary state through relaxation in the stable regime. In subsequent evolution the system is very close to this state which makes the oscillations unobservable. In contrast, the spiked pulsed emission of Fig.~\ref{fig:pulsed}(a) can be understood as the result of a large deviation of the initial state from the stationary state in this case. The oscillations in parameter space are so large in amplitude that they no longer have a harmonic character but are characterized by periods of complete disappearance of the condensate density, as shown in Fig.~\ref{fig:phasespace}(b).

The above considerations indicate that the range in parameter space which correspond to oscillations is mostly placed in the short polariton lifetime region. Similar condition has been previously determined for the polariton condensate instability in the case of continuous nonresonant pumping~\cite{Natalia_1}. Both oscillations and instability appear to be related to the nonadiabaticity of the evolution, when the active reservoir and condensate density cannot be reduced to a single degree of freedom~\cite{Natalia_adiabatic}. However, we note that the two phenomena are very different, which is clear form the fact that the instability relies on the polariton-reservoir interactions, while in the description of relaxation oscillations the effect of interactions can be neglected on the level of Eqs.~(\ref{GP_single}).

\section{\label{sec:CONCLUSION}CONCLUSION}

In conclusion, we studied relaxation oscillations in an exciton-polariton condensate analytically and numerically using a model including both active and inactive reservoir. We derived a second order differential equation for condensate density, applicable in the regime of continuous wave excitation or pulsed excitation provided that inactive reservoir is decaying slowly. We investigated the properties of fixed points of the equation and concluded that oscillations can occur only around the nontrivial second fixed point which appears in the nonlinear regime. 
We derived an analytical condition for the appearance of oscillations consistent both with numerical simulations and with previous experimental observations. We demonstrated that oscillatory behaviour should be observable mainly in low quality microcavities, i.e. having short photon lifemes. Additionally, we presented regimes of oscillating and spiked pulsating dynamics, and described the corresponding trajectories in the phase space. {\color{red}}

\acknowledgments
We acknowledge support from the National Science Centre, Poland, grants
2015/17/B/ST3/02273, 2016/22/E/ST3/00045, and 2016/23/N/ST3/01350.

\appendix*
\section{}
Below, we present the parameters used in the numerical simulations, summarized in Table~1.

\begin{table}[h]
\caption{\label{tab:table1} The parameters  used  in  simulations  to  obtain  the  results
presented in the work. 
In addition, the parameters used in the simulation are  \(g_C\)=3.4 \(\mu\)eV\(\mu m^2\), \(g_R\)=\(g_I\)=2\(g_C\), \(m^*_{LP}=3\cdot 10^{-5} m_e\).}
\begin{ruledtabular}
\begin{tabular}{ c c c c c}
Figures & Fig.1  & Fig.2(a) & Fig.4(a), Fig.5&\\
\\
\hline
$\tau_C $           &  2                    &  0.5                  & 1                             &$ps$                  \\
$\tau_R$            & 40                    & 800                   & 100                           &$ps$                  \\
$\tau_I$            & 1000                  & 2000                  & 1000                          &$ps$                  \\
$R$        & $8 \cdot 10^{-3}$       & $10^{-2}$               & $10^{-3}$                       &$\frac{\mu m^2}{ps}$  \\
$\kappa^{1D}$   &$ 5 \cdot 10^{-2}$       &  $10^{-5}  $            &$ 5 \cdot 10^{-2} $              &$\frac{\mu m^2}{ps}$  \\
$n_I^{0}$       & -                     & 1800                  & -                             &$\frac{1}{\mu m^2}$   \\
$ P $             & 1.25                  & -                     & 3                             &$P_{th}$              \\

\end{tabular}
\end{ruledtabular}
\end{table}  
\newpage
\bibliography{references_PRB}
\end{document}